\documentclass[twocolumn,showpacs,preprintnumbers,amsmath,amssymb,nofootinbib]{revtex4}
\input psfig.sty
\usepackage{graphicx}% Include figurthebibliographye files
\usepackage{dcolumn}% Align table columns on decimal point
\usepackage{bm}% bold math

\begin{document}

\def\be{\begin{equation}}
\def\ee{\end{equation}}
\def\lb{\label}

%\preprint{APS/123-QED}

\title{Nonextensive effects on the relativistic nuclear equation of state}

%\title{The Walecka Theory in the Framework of the Nonextensive Statistical Mechanics}

%%%%%%%%%%%%%%%%%%%%%%%%%%%%%%%%%
\author{F. I. M. Pereira$^{1}$} \email{flavio@on.br}
\author{R. Silva$^{2}$} \email{raimundosilva@uern.br}
\author{J. S. Alcaniz$^{1}$} \email{alcaniz@on.br}
%\author{C. A. Z. Vasconcellos$^{3}$} \email{cesarzen@if.ufrgs.br}

\affiliation{$^{1}$Observat\'orio Nacional, Rua Gal. Jos\'e Cristino 77, 20921-400 Rio de Janeiro RJ, Brasil}
\affiliation{$^{2}$Universidade do Estado do Rio Grande do Norte, 59610-210, Mossor\'o, RN, Brasil}
%\affiliation{$^{3}$Instituto de F\'\i sica, Universidade Federal do Rio
%Grande do Sul, Porto Alegre, RS, Brasil}

\date{\today}% It is always \today, today,
             %  but any date may be explicitly specified

\begin{abstract}
The Walecka many-body field theory is investigated in the context of  quantum nonextensive statistical mechanics, characterized by a dimensionless parameter $q$. In this paper, we consider nuclear matter described statistically by a power-law distribution which generalizes the standard Fermi-Dirac distribution ($q = 1$).  We show that  the scalar and vector meson fields become more intense due to the nonextensive effects ($q \neq 1$). From a numerical treatment, we also show that as the nonextensive parameter $q$ increases, the nucleon effective mass diminishes and the equation of state becomes stiffer. Finally, the usual Maxwell construction seems not to be necessary for isotherms with temperatures in the range $14 {\rm{Mev}} < k_BT < {\rm{20 MeV}}$.
\end{abstract}

% PACS,
                             % Classification Scheme.
%\keywords{Suggested keywords}%Use showkeys class option if keyword
                              %display SL99desired
\maketitle

\section{Introduction}

In the past few years, considerable attention has been paid to the so-called nonextensive statistics, both from theoretical and observational viewpoints. The nonextensive framework, first developed by Tsallis \cite{T88,SL99}, seems to present a consistent theoretical tool to investigate complex systems in their
nonequilibrium stationary states. Recently, several consequences (in different branches) of the Tsallis framework have been investigated in the literature \cite{todos}, which includes systems of interest in high energy physics\footnote{An updated bibliography on Tsallis' nonextensive statistics can be found at http://tsallis.cat.cbpf.br/biblio.htm}. In this regard, the very first application was addressed to a problem of solar neutrino \cite{kaniadakis96}. In this study, the authors applied the $q$-framework to derive a distribution function for the interior plasma, which is relevant in the calculation  of the nuclear fusion reaction rates responsible for the neutrino flux emitted by Sun. Recently, within Fokker-Planck dynamics, the Einstein's relation between drag, diffusion, and the equilibrium distribution for a spatially
homogeneous system have been generalized in the context of Tsallis statistics, and such a model was applied to charm quark dynamics in a thermal quark-gluon plasma for the case of collisional equilibration \cite{walton00}. More recently, interpretations for central Au-Au collisions at RHIC energies in a Relativistic
Diffusion Model (RDM) for multiparticle interactions based on the interplay of nonequilibrium and local equilibrium were studied also through extensive and nonextensive statistics \cite{george04}.

In the theoretical treatment of the properties of nuclear matter, the relativistic phenomenological approach developed by Walecka \cite{SW} (the so-called quantum hadrodynamics [QHD-I]), represents one of the important approaches to the highly nonlinear behavior of strong interactions at the hadronic energy scales. This model provides a thermodynamically consistent theoretical framework for the description of bulk static properties of strong interacting many-body nuclear systems\footnote{As is well known, although providing a consistent theoretical framework, QHD-I presents some limitations.  In this regard, a more complete theoretical treatment (for zero temperature) was discussed in Ref.~\cite{CFH}, whereas the thermo field dynamics in hot nuclear matter was considered in Ref.~\cite{Hen}. %Such important treatments lie, however, out the scope of the present work. 
Here, however, we limit ourselves to the effects of nonextensivity on the QHD-I theory, as originally developed in Ref.~\cite{SW}.}.

In this work we investigate the effects of the nonextensive statistical mehanics on the QHD-I theory by considering a $q$-generalization of the Fermi-Dirac (FD) distributions. In this regard, it is worth emphazing that earlier generalizations of quantum statistics for fermions and
bosons have been developed in Ref. \cite{DBG} and applied in the context of relativistic nuclear equation of state (EoS) of Boguta-Bodmer \cite{BB} by Drago et al \cite{Drago}. However, in the present work, to study the nonextensive effects on the QHD-I theory we use  the most
recent fermion distribution obtained by Tweldeberhan, Plastino and Miller in Ref. \cite{TPM} (TPM distribution). This  development  differs from the above cited in that it provides a new cut-off prescription based on the extremization of a thermodynamical functional. This paper is organized as follows. In Sec. II, we present the basic formalism of the mean field theory of QHD-I for the calculation of some important quantities of nuclear matter. A brief review of Tsallis statistics and TPM distribution function is made in Sec. III. In Sec. IV the convergence of the calculation is also considered and it is shown that the allowed values of $q$ lie in the range $1<q<5/4$. Our main results are discussed in Sec. V. We summarize the main conclusions in Sec. VI.

\section{Basics of QHD-I }

As widely known, the Lagrangian density describing the nuclear matter reads \cite{SW}
\begin{eqnarray}  \label{DensLagran1}
{\cal L} =\bar{\psi }[(i\gamma _\mu (\partial ^\mu
-g_\omega\omega^{\mu})-(M-g_\sigma
\sigma)]\psi \nonumber \\
 + \frac 12(\partial _{\mu} \sigma \partial ^{\mu}
\sigma-m_{\sigma}^{2} \sigma^{2})-\frac {1}{4}\omega_{\mu \nu}
\omega^{\mu \nu}
 + \frac{1}{2}m_{\omega}^{2}\omega_{\mu} \omega^{\mu}~,
\end{eqnarray}
which represents nuclear matter composed by nucleons coupled to two mesons, namely, the $\sigma$ and $\omega$ mesons (for details see reference \cite{SW}).

Applying standard techniques from field theory  and the mean-field approach, we obtain the scalar density
\begin{equation}
\varrho_S=\frac{\gamma_{\rm N}}{(2\pi)^3}\int\frac{M^*}{E^*(k)}[n(\nu,T)+{\bar{n}}(\nu,T)]d^{3}k~,
\label{RS}
\end{equation}
where $M^*$ is the effective mass
\begin{equation}
M^*=M - g_{\sigma}\sigma = M-\frac{g_\sigma^2}{m_\sigma^2}\rho_S\;.
\label{ms}
\end{equation}
The baryon number density, the energy density and pressure are given, respectively, by
\begin{equation}\label{RB}
\varrho_B=\frac{\gamma_{\rm
N}}{(2\pi)^3}\int[n(\nu,T)-{\bar{n}}(\nu,T)]d^{3}k,
\end{equation}
\begin{eqnarray}
\varepsilon&=&\frac{1}{2}\frac{g_\omega^2}{m_\omega^2}\varrho_B^2
+\frac{1}{2}\frac{g_\sigma^2}{m_\sigma^2}(M-M^*)^2+\nonumber\\
&&\frac{\gamma_{\rm N}}{(2\pi)^3}\int{E^*(k)}
[n(\nu,T)+{\bar{n}}(\nu,T)]d^{3}k\;,
\end{eqnarray}
\begin{eqnarray}\label{press}
p&=&-\frac{1}{2}\frac{g_\omega^2}{m_\omega^2}\varrho_B^2
+\frac{1}{2}\frac{g_\sigma^2}{m_\sigma^2}(M-M^*)^2+\nonumber\\
&&\frac{1}{3}\frac{\gamma_{\rm N}}{(2\pi)^3}\int\frac{k^{2}}{E^*(k)}
[n(\nu,T)+{\bar{n}}(\nu,T)]d^{3}k\;,
\end{eqnarray}
where
\begin{equation}
\label{FD}
 n(\nu,T)=\frac{1}{{\rm
e}^{\beta[E^*(k)-\nu]}+1}
\end{equation}
and
\begin{equation}
\bar{n}(\nu,T)=\frac{1}{{\rm e}^{\beta[E^*(k)+\nu]}+1},
\end{equation}
are the usual FD distributions for baryons and anti-baryons, with $E^*(k)=\sqrt{k^2+{M^*}^2}$, $\beta=1/k_BT$. The parameter
$\nu\equiv\mu-g_\omega\omega_0=\mu-(g_\omega/m_\omega)^2\varrho_B$
is the effective chemical potential, and $\gamma_N$ is the
multiplicity factor ($\gamma_N=2$ for pure neutron matter and
$\gamma_N=4$ for nuclear matter).

 Additionally, we also use for the coupling constants the values of reference
\cite{SW}, namely\footnote{For the purpose of the present work, the values given in Eq. (\ref{cpcts}) suffices to investigate the effects of the nonextensivity in neutron and nuclear matter.  Variations of the coupling constants, within the acceptable
values given in current literature, do not qualitatively affect the conclusions.},
\begin{equation}
\label{cpcts}
\bigg(\frac{g_\sigma}{m_\sigma}\bigg)^2=11.798~{\rm fm^2}~{\rm~and}~
\bigg(\frac{g_\omega}{m_\omega}\bigg)^2=8.653~{\rm fm^2}~,
\end{equation}
which are fixed to give the bind energy $E_{\rm bind}=-15.75$ MeV and $k_F=1.42$ $\rm{fm}^{-1}$.

\section{Non-extensive framework }

Over the past two decades, nonextensive statistical mechanics has
successfully addressed a wide range of nonequilibrium phenomena in
non-ergodic and other complex systems [1, 2]. As widely known,
nonextensive statistical mechanics, as proposed by Tsallis [1], is a
generalization of Boltzmann-Gibbs (BG) statistical mechanics based
on the functional
\begin{equation}
S_{q} = -k_B \sum_{i=1}^W p_i^q \ln_q p_i\;,\quad
S_{q = 1}=-k_B\sum_{i=1}^{W}p_i\ln p_i.
\end{equation}
%Here and hereafter, the Boltzmann constant is set equal to unity forthe sake of simplicity, 
%%%%%%%%%%%%%%%%%%%
Here, $q$ is the nonextensive parameter and the $q$-logarithmic function above is defined as
\begin{equation}
\ln_q p ={p^{1-q}-1\over 1-q}.
\end{equation}

In this $q$-framework, the additivity for two probabilistically
independent subsystems A and B is generalized by the following
pseudoadditivity
\begin{equation}\label{S-q}
{S^{A U B}_{q}}= {S_{q}^{A}} + {S^{B}_{q}}+(1-q){S_{q}^{A}}
{S^{B}_{q}}\,\,\,,
\end{equation}
where the cases $q < 1$ and $q > 1$ correspond to super-additivity and
sub-additivity, respectively. For subsystems that have special
probability correlations, extensivity is not valid for
Boltzmann-Gibbs entropy, but may occur for $S_q$ with a particular
value of the index $q \neq 1$. Such systems are sometimes referred
to as nonextensive \cite{tsallis05}.

\begin{figure*}[t]
\vspace{.2in}
\centerline{\psfig{figure=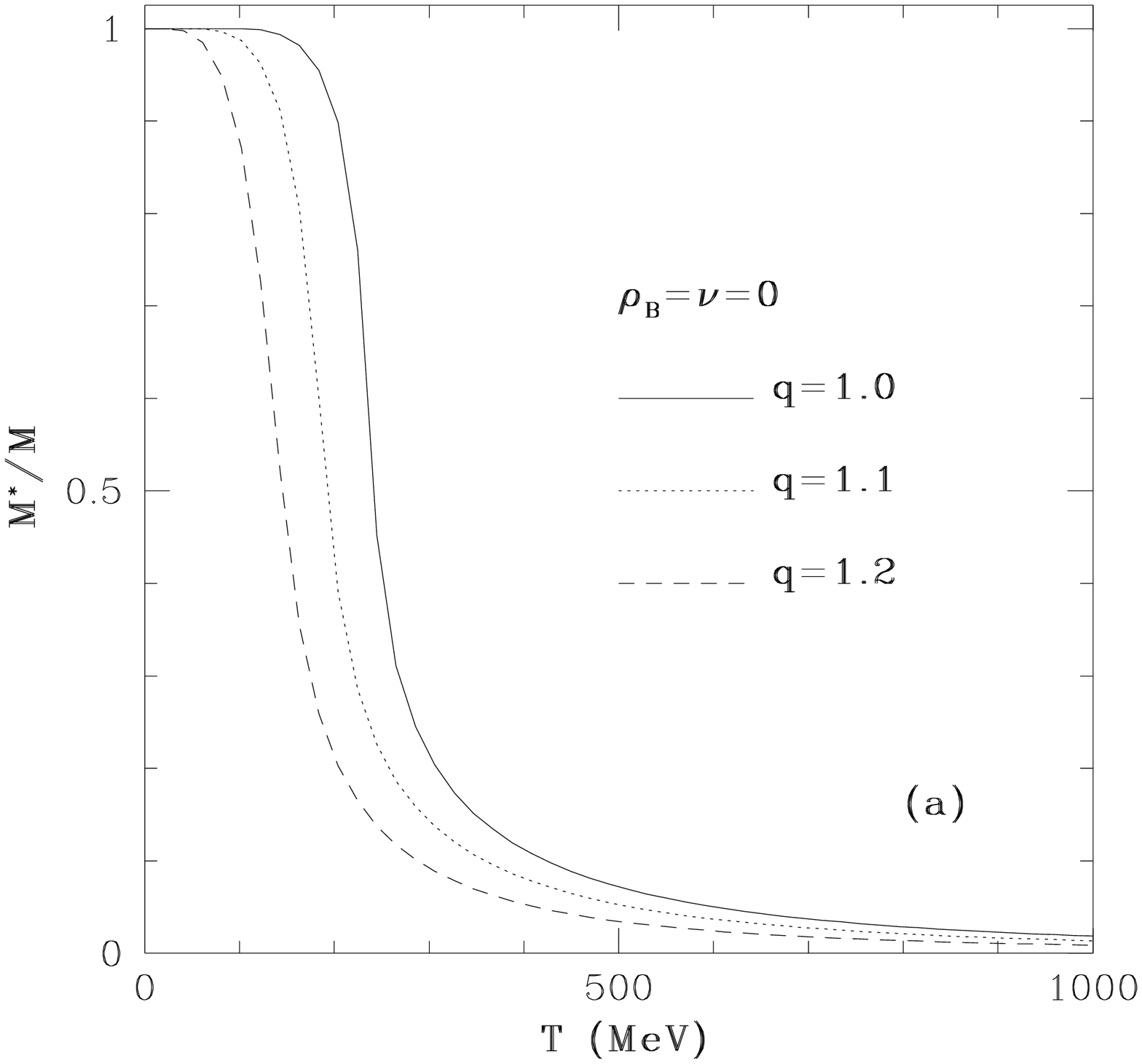,width=3.3truein,height=3.0truein}\hskip
.25in \psfig{figure=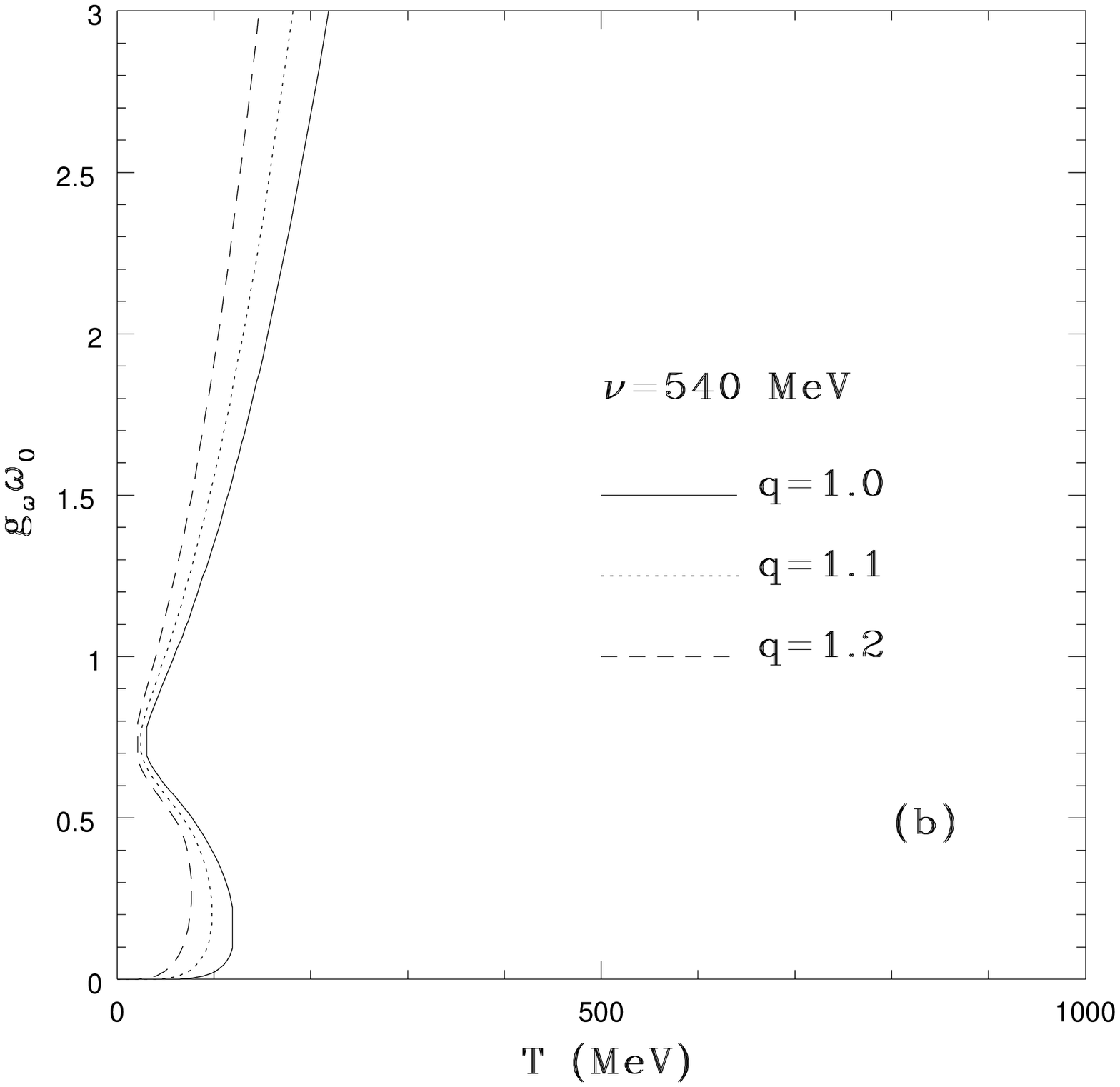,width=3.3truein,height=3.0truein}
\hskip .5in} \caption{The effective nucleon mass and the vector
meson field of pure neutron matter ($\gamma_N=2$) as function of
temperature for different values of the parameter $q$. Panel (a):
the self-consistent nucleon mass at vanishing baryon density. Panel
(b): the vector meson field at nonzero baryon density corresponding
to $\nu=540$ MeV.} \label{fig1}
\end{figure*}

\subsection{Quantum Statistics}

We recall the main aspects of the connections between the quantum statistics and Tsallis framework. Specifically, we concentrate on the new cut-off prescription related to the Tsallis' maximum entropy distributions (for details see Ref. \cite{TPM}). The main result in this study is that, for $q>1$, a $q$-generalized
quantum distributions for fermions and bosons are given by
\begin{equation}\label{nq}
 n_q(\mu,T)=\frac{1}{\tilde{e}_q(\beta(\epsilon-\mu))\pm1},
  \end{equation}
where $e_q$ reads
\begin{eqnarray}
\label{TPM}\tilde{e}_q(x)= \left\{
\begin{array}{l}
~[1+(q-1)x]^{\frac{1}{q-1}}~~~~~{\rm if}~~~~~~x>0~~~~~~~~~~~~\\
\\
~[1+(1-q)x]^{\frac{1}{1-q}}~~~~~{\rm if}~~~~~~x\leq0~.~~~~~~~~~\\
\end{array}
\right.\
\end{eqnarray}
and $x=\beta(\epsilon-\mu)$. In the $q\rightarrow1$ limit, the standard FD distribution, $n(\mu,T)$, is recovered. As physically expected,  as $T\rightarrow0$, $n_q(\mu,T)\rightarrow n(\mu,T)$. This amounts to saying that for studies of the interior of neutron stars (where, in nuclear scale, $T \simeq 0$) we do not expect any nonextensive signature. On the other hand, in heavy ions collision experiments or in the interior of protoneutron stars, with typical stellar temperatures of several tens of MeV (1 MeV$=1.1065\times10^{10}$ K), nonextensive effects may appear. In order to study such effects, in the next Section we combine Eqs.(\ref{ms})-(\ref{press}) with the generalized FD distributions given by Eqs. (\ref{nq}) and (\ref{TPM}).

\begin{figure*}[t]
\vspace{.2in}
\centerline{\psfig{figure=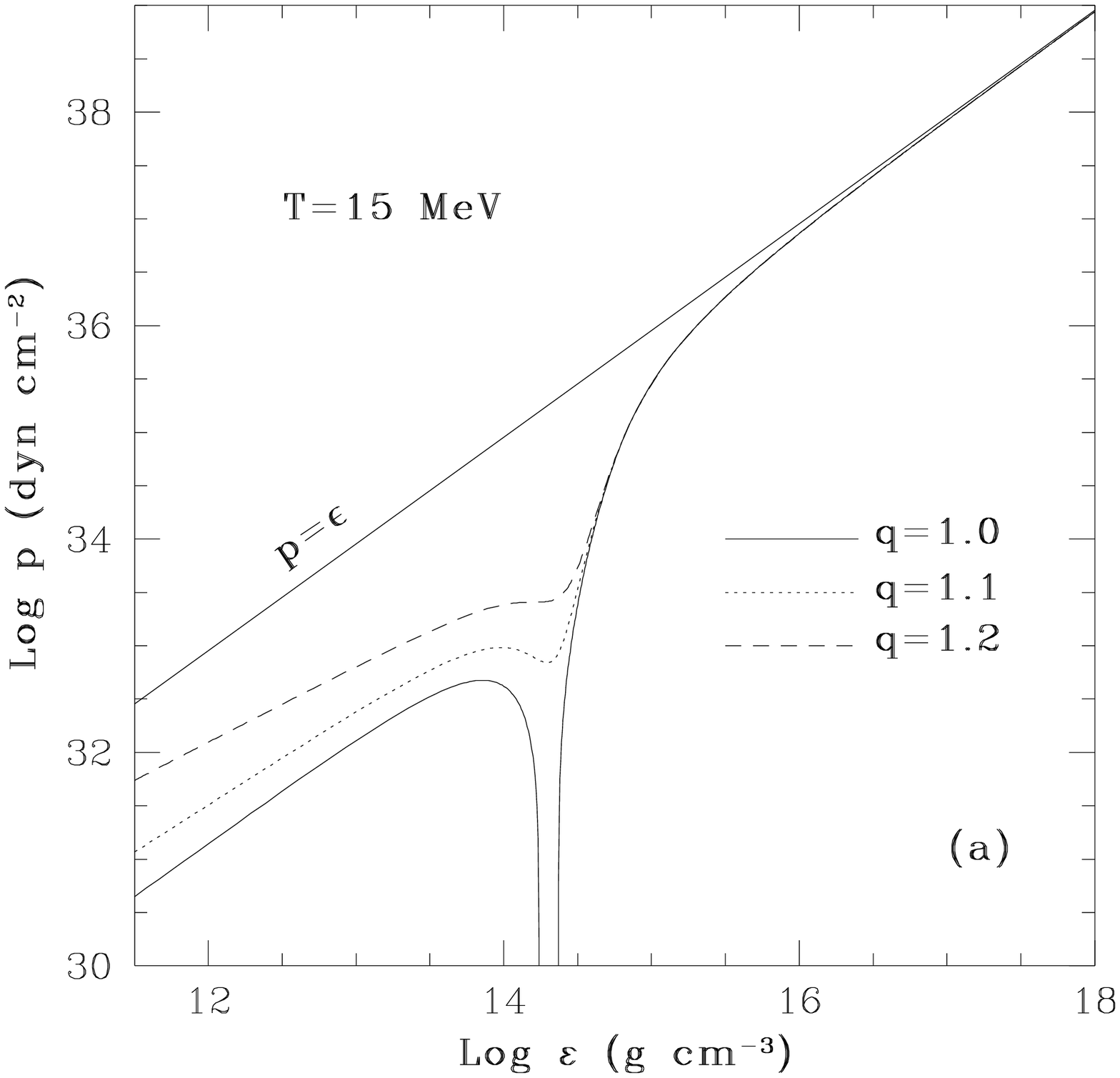,width=2.5truein,height=3.0truein}
\psfig{figure=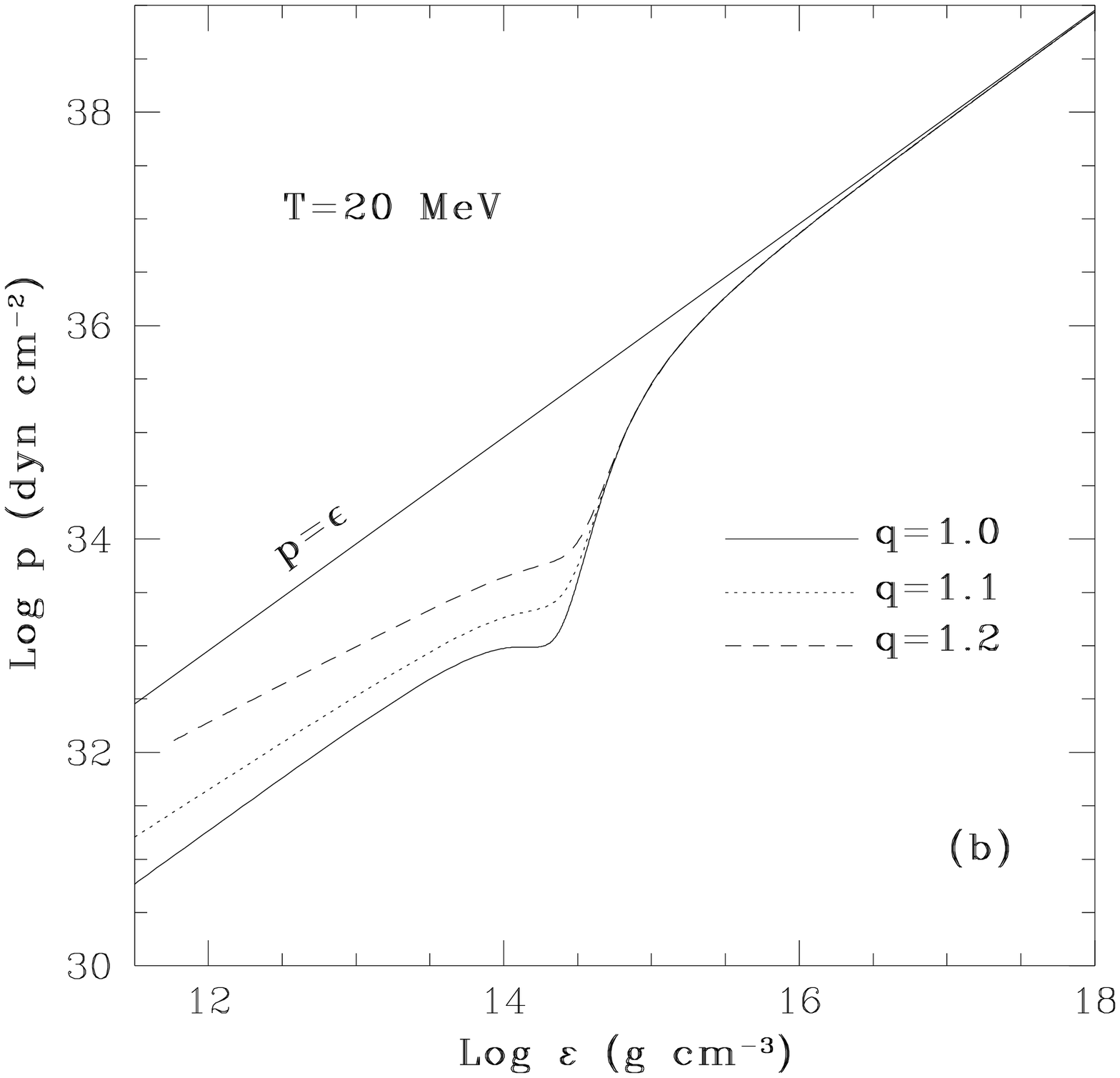,width=2.5truein,height=3.0truein}
\psfig{figure=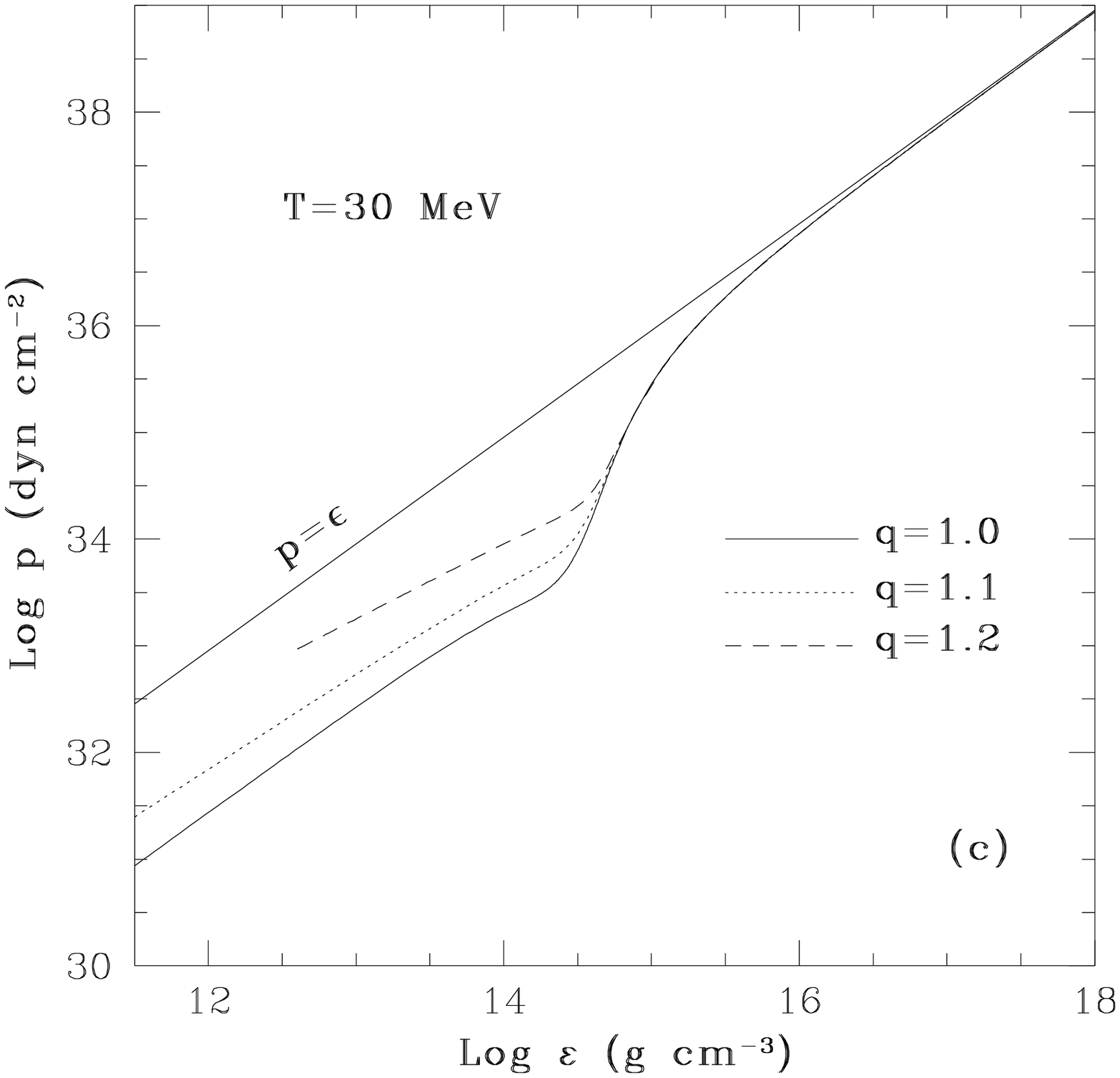,width=2.5truein,height=3.0truein}\hskip .5in}
\caption{ Isotherms of nuclear matter ($\gamma_N=4$) equation of
state at finite temperatures for different values of the parameter
$q$.} \label{fig2}
\end{figure*}

\section{Nonextensivity and QHD-I}

Since the above function $\tilde{e}_q(x)$ is a \emph{deformed} exponential, we must verify the mathematical convergence of the
integrals in Eqs. (\ref{ms})-(\ref{press}) by considering the $q$-distributions. Thus, let us write the  $\tilde{e}_q(x)$ in a more
compact form given as
\begin{equation}
\tilde{e}_q(x)\equiv(1+\xi x)^{1/\xi}
\label{exi}
\end{equation}
where $\xi\equiv q-1$. Now, taking into account that
\begin{equation}
\tilde{e}_q(x)\longrightarrow \xi^{1/\xi} x^{1/\xi},
\label{exi1}
\end{equation}
in the limit $x>>1$, we obtain the asymptotic behaviour for the integrals appearing in Eqs.(\ref{ms})-(\ref{press}), i.e.,
\begin{equation}
\label{int1}
M^*:\int\frac{M^*~d^3k}{E^*(k)\{\tilde{e}_q[E^*(k)-\nu]+1\}}
\longrightarrow \frac{k^2}{\xi^{1/\xi}~k^{1/\xi}},\\
\end{equation}
\begin{equation}
\rho_B:\int\frac{d^3k}{\tilde{e}_q[E^*(k)-\nu]+1}
\longrightarrow\frac{k^3}{\xi^{1/\xi}k^{1/\xi}},\\ \label{int2}
\end{equation}
\begin{equation}
\varepsilon:\int\frac{E^*(k)~d^3k}{\tilde{e}_q[E^*(k)-\nu]+1}
\longrightarrow\frac{k^4}{\xi^{1/\xi}k^{1/\xi}},\\ \label{int3}
\end{equation}
\begin{equation}
p:\int\frac{k^2~d^3k}{E^*(k)\{\tilde{e}_q[E^*(k)-\nu]+1]\}}
\longrightarrow \frac{k^4}{\xi^{1/\xi}~k^{1/\xi}}~.
\label{int4}
\end{equation}

From Eqs. (\ref{int1})-(\ref{int4}) the general asymptotic behaviour can be summarized by ${k^N}/{k^{1/\xi}}$, for $N=2, 3, 4$.
In order to have ${k^N}/{k^{1/\xi}}\longrightarrow0$, when $k\longrightarrow\infty$, we find that ${1}/{\xi}>N$, from which
we obtain
\begin{equation}
~~q<\frac{N+1}{N}~~~~({\rm~for}~~q>1)~. \label{qM1}
\end{equation}
Note that, to simultaneously satisfy the convergence of all integrals in Eqs.(\ref{ms})-(\ref{press}), we find that $1<q<5/4$, consistent with the limits of $q$
in the range $q\in(0, 2]$ of Ref. \cite{AR}.

At high temperatures ($T\rightarrow\infty$), the analytic solution to Eq. (\ref{ms}) can be written as 
\be
M^*\rightarrow M\bigg[1+\frac{g_\sigma^2}{m_\sigma^2}
\bigg(\frac{\gamma_N}{\pi^2}\bigg)\xi_1(q)(k_BT)^2\bigg]^{-1}
\lb{Mss}
\ee
where $\xi_1(q)$ is given by Eq. (\ref{xin}).

Several limiting cases of the EoS are of interest:

\begin{enumerate}

\item The baryon distribution becomes a step function
$n_q(\nu,0)=\theta(k_F-|{\bf k}|)$ in the limit $T\rightarrow0$ for any value of $\rho_B$.

\item The system becomes degenerate in the limit $\rho_B\rightarrow\infty$ at any $T$.

\item For $k_BT<<M$ and $\rho_B\rightarrow0$, the equation of state of a classical
nonrelativistic gas is obtained:

\be
\varepsilon=\frac{2}{7-5q}[(2-q)M\rho_B+\frac{3}{2}\rho_Bk_BT]
\lb{Eclass}
\ee
and
\be
p=\frac{2}{7-5q}[(q-1)M\rho_B+\rho_Bk_BT]~.
\lb{Pclass}
\ee
 Thus, the equation of state of a nonrelativistic gas of baryons can also be given by
\be
p=\bigg[\frac{2(q-1)M+2k_BT}{2(2-q)M+3k_BT}\bigg]~\varepsilon~.
\lb{EoSclass}
\ee
Note that, as $q\rightarrow1$, the limits of Ref. \cite{SW} are recovered.

\item As $T\rightarrow\infty$ for any vaue of $\rho_B$, an equation of state similar to that of a black
body is obtained:
\be
\varepsilon\rightarrow\frac{\gamma_N}{\pi^2}\xi_3(q)(k_B T)^4~,~~~p=\varepsilon/3~,
\lb{Eq}
\ee
where, in Eqs.(\ref{Mss}) and (\ref{Eq}),
\be
\xi_n(q)\equiv\int_0^\infty\frac{z^n~dz}{\tilde{e}_q(z)+1}~.
\lb{xin}
\ee
When $q\rightarrow1$ we have that $\xi_1(q\rightarrow1)\rightarrow\pi^2/12$
and $\xi_3(q\rightarrow1)\rightarrow7\pi^4/120$, recovering the limits of Ref. \cite{SW}.
\end{enumerate}

\section{Results}

\begin{figure*}[tbh]
\vspace{.2in}
\centerline{\psfig{figure=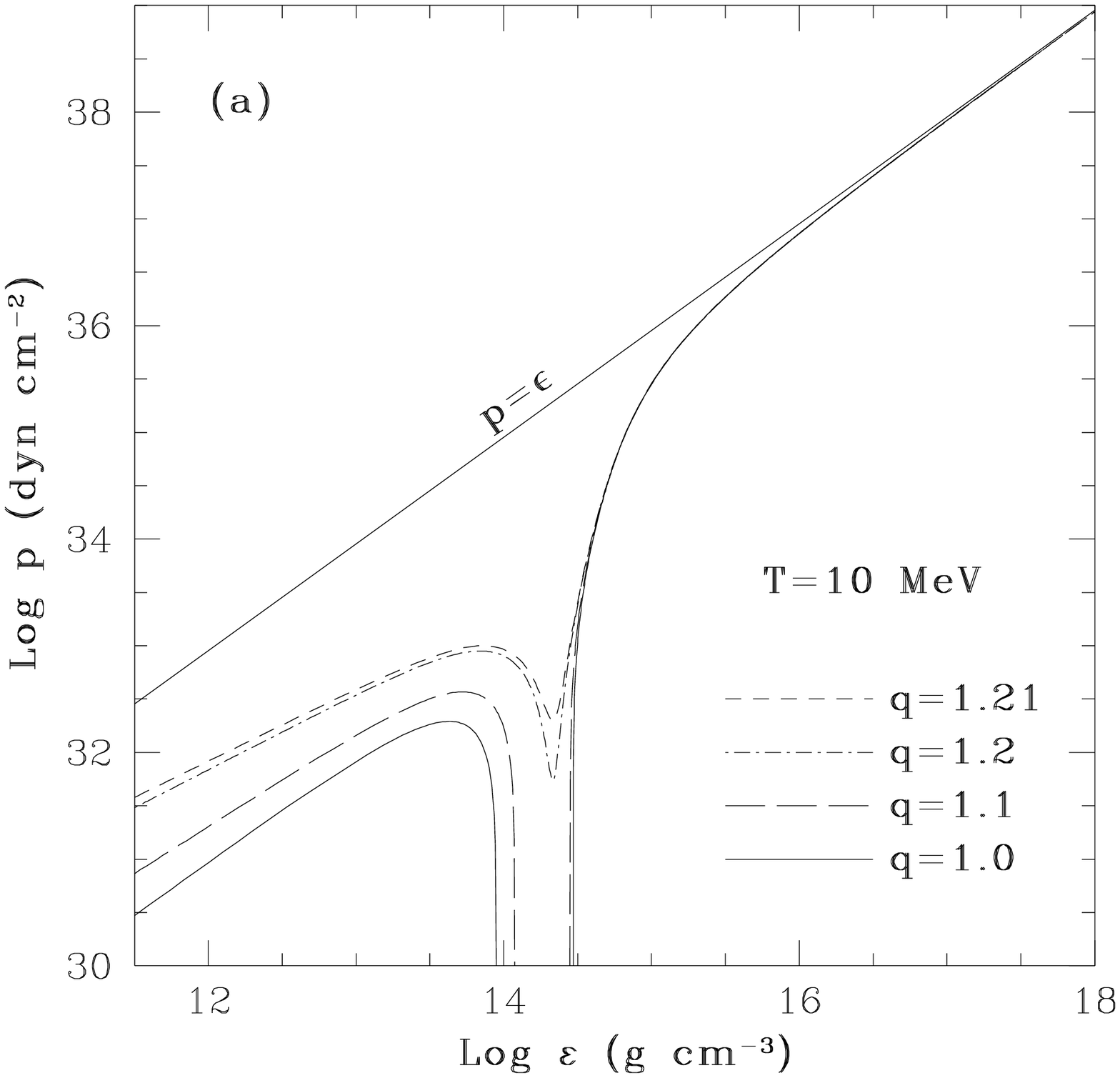,width=2.5truein,height=3.0truein}
\psfig{figure=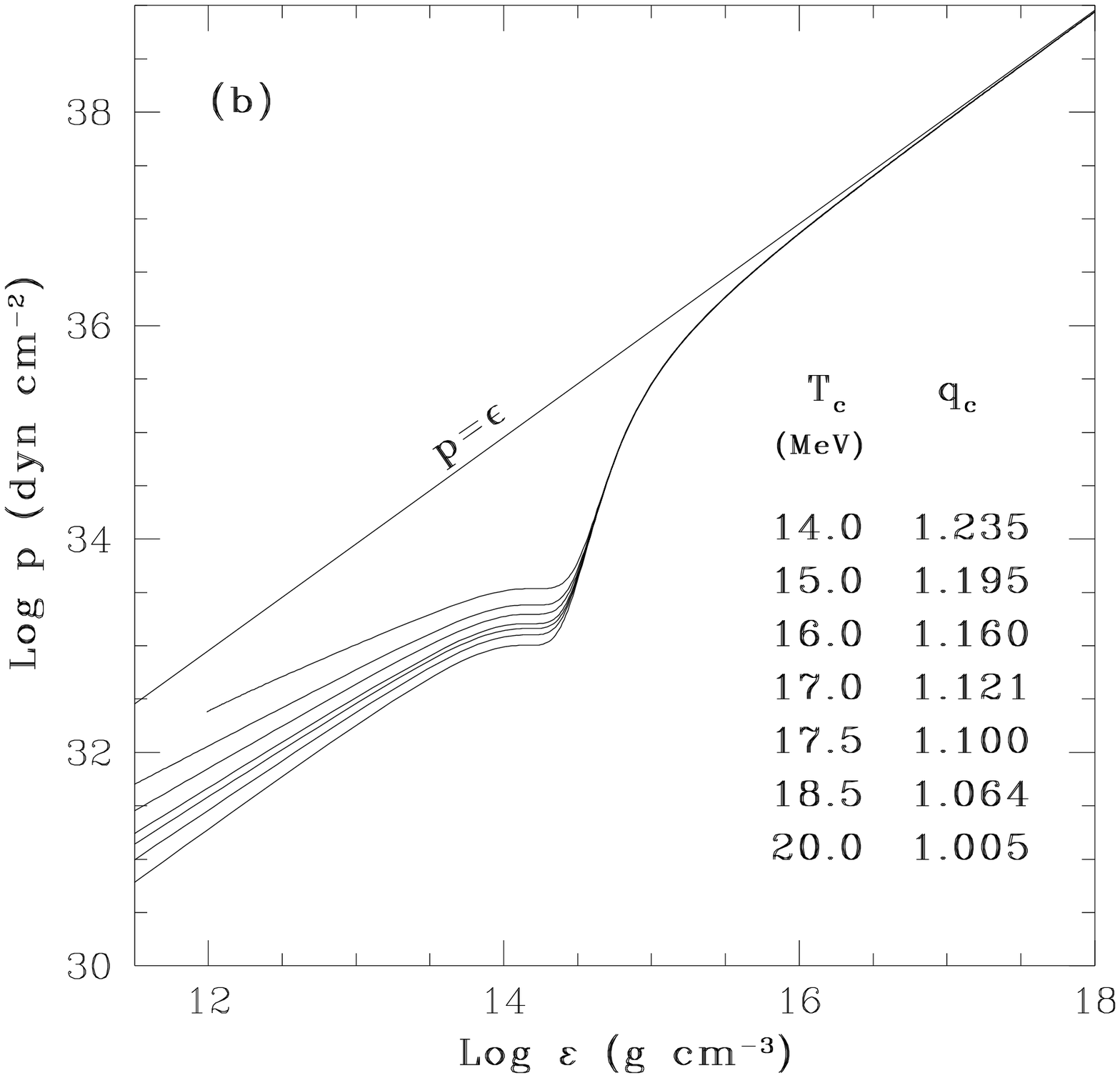,width=2.5truein,height=3.0truein}
\psfig{figure=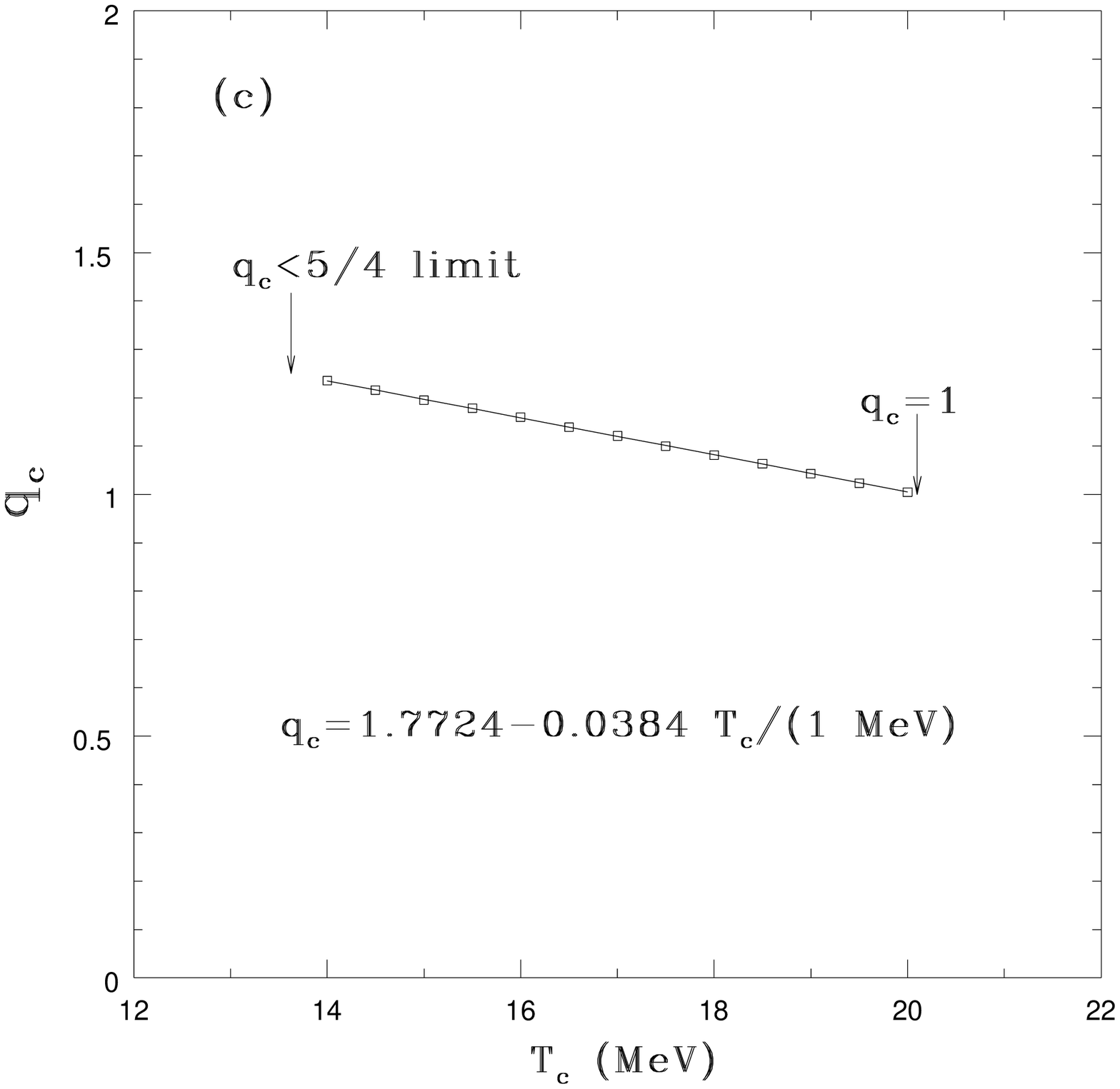,width=2.5truein,height=3.0truein}\hskip .5in}
\caption{Panel (a): Isotherms of nuclear matter ($\gamma_N=4$) at
$T=10$ MeV for several values of the parameter $q$. Panel (b):
Isotherms of nuclear matter corresponding to the critical
temperatures and parameters $q_c$ given in the data. The curves and
the data are in the same order from top to bottom. Panel (c): the
critical parameter $q_c$ as funtion of the critical temperature
$T_c$ corresponding to the data of Panel (b).} \label{fig3}
\end{figure*}

In order to study the nonextensive effects on QHD-I theory, we have
calculated numerically, for several values of temperatures and of
the parameter $q$, the effective nucleon mass, the vector and scalar
mesons fields for pure neutron matter ($\gamma_N=2$), as well as the
EoS for symmetric nuclear matter ($\gamma_N=4$).

In Fig. \ref{fig1} we show the nonextensive effects on the effective mass and vector meson for a pure neutron matter. In Panel (a), for $\varrho_B= \nu =0$, we note that the higher the parameter $q$,  the smaller the effective mass $M^*$ (and, consequently, the higher the scalar field $\sigma$, since $g_\sigma\sigma=M - M^*$). Such an effect may be physically understood in that at a given temperature, the scalar density, as source for the scalar mesons, increases with the increasing of the nonextensive parameter $q$. Thus, the attraction of the nucleons, mediated by the scalar mesons, becomes stronger, reducing the effective mass. The behaviour of vector meson field $g_\omega\omega_0=(g_\omega/m_\omega)^2\varrho_B$ is shown in Panel (b).

Fig. \ref{fig2} shows the nonextensive effect on the symetric nuclear matter EoS. The panels display the $\log{p}-\log{\epsilon}$ plane for selected values of $q$. The results are ploted for arbitrarily chosen values of temperature, $T=$15 MeV, 20 MeV, and 30 MeV.  In reality, the motivation for this choice is of astrophysical interest in the study of protoneutron stars. Clearly, the nonextensive effect is manifested in the increasing of the pressure with the values of $q$ becoming the EoS stiffer.

Another interesting effect of the nonextensivity on nuclear EoS of QHD-I concerns the phase transitions. From Panels (a) and (b) of Fig. \ref{fig2}, we see that the
first order phase transition may be eliminated by the variation of the parameter $q$. This fact can be easily visualized in Fig. \ref{fig3}{(a)}, where isotherms at $T=10$ MeV are plotted for several values of $q\in[1.0,\;5/4)$. Note that, for increasing values of $q$, the dip in the region of thermodynamical
instability becomes smaller, vanishing at the turning point that defines the critical values of
thermodynamical quantities ($T_c$, $p_c$, etc.). %One may conclude, therefore, that the order of the phase transition depends on the value of the nonextensive %parameter $q$.
We also note that for our choice $T=10$ MeV the upper value of $q$
(near the 5/4 limit discussed earlier) is not sufficient to
eliminate the first order phase transition \cite{text}. On the other hand, the phase transitions can be eliminated for values of $T$ in the
region $14{\rm MeV}<T<20{\rm MeV}$. In this interval, all
temperatures can be made critical. This amounts to say that a
(critical) parameter $q_c$ can be determined in order to yield a
turning point in the isotherm at a given temperature. This is
illustrated in Fig. \ref{fig3}(b) where several isotherms are
displayed. In Fig. \ref{fig3}(c)  values of $q_c$ are plotted as
function of temperature in the above range ($14~{\rm MeV}<T<20~{\rm MeV}$).
The smooth behavior of $q_c$ allows a parametrization of the form
\begin{equation} 
\label{qc}
q_c=1.7724(\pm0.003)-0.0384(\pm0.0001) \frac{k_BT}{(1{\rm
MeV})} \;.
\end{equation}

\section{final remarks}

In this paper, we have investigated the nonextensive effects on the mean field theory of Walecka (QHD-I) \cite{SW}.  We have used, instead of the standard Fermi-Dirac nucleon and antinucleon distribution functions, the $q$-quantum distribution recently obtained by Teweldeberhan, Plastino and Miller in Ref. \cite{TPM},  which has a new cut-off prescription based on the extremization of a thermodynamical functional. We emphasize that the nonextensive effects on nuclear and pure neutron matter, for a considerable range of temperature, is to make the equation of state stiffer and to increase the intensity of the vector and scalar meson fields, with a consequent lowering of the nucleon effective mass (for increasing values of the nonextensive parameter $q$). We believe that it may have interesting consequencies in astrophysical studies, mainly in what concerns the calculation of masses of compact objects, such as protoneutron stars.

As discussed in Sec. V, another interesting feature of nonextensivity is that, at temperatures in the range $14~{\rm MeV}<T<20~{\rm MeV}$, phase transitions of first order can be avoided by a convenient variation of the parameter $q$, which allows the determination of a critical $q_c$ parameter at the turning point of an isotherm at a given $T$ [Eq. (\ref{qc})]. Finally, it is worth mentioning that the estimates for the nonextensive parameter from the \emph{generalized} Walecka many-body field theory is consistent with the upper limit $q < 2$, obtained from several independent studies in the quantum \cite{AR} and in the non quantum limit \cite{20} involving the Tsallis nonextensive
framework.

{\bf Acknowledgments:} The authors are very grateful to C. A. Z. Vasconcellos for helpful discussions. RS and JSA are  partially supported by the Conselho Nacional de Desenvolvimento Cient\'{\i}fico e Tecnol\'{o}gico (CNPq - Brazil).
JSA is also supported by FAPERJ No. E-26/171.251/2004.


\begin{thebibliography}{30}
\bibitem{text} At this limit or above it the convergence of the integrals in Eqs. (\ref{ms})-(\ref{press}) breaks down, being necessary the use of the Maxwell construction. It is worth emphasizing that this picture comes from QHD-I so that it can be very different in others theories.
\end{thebibliography}
\end{document}